\documentclass[11pt]{article}
\usepackage[affil-sl,auth-sc]{authblk}
\usepackage{amssymb,amsmath,amsbsy}
\usepackage{mathrsfs}
\usepackage{mathbbol}
\usepackage{dsfont}
\usepackage{appendix}
\usepackage{hyperref}
\usepackage[top=1.2in, bottom=1.2in,left=0.9in,includefoot]{geometry}               % See geometry.pdf to learn the layout options. There are lots.
\usepackage{graphicx}
\usepackage{cite}
\usepackage[usenames,dvipsnames]{color}
\def\theequation{\arabic{section}.\arabic{equation}}
\def\eq#1{eq.~(\ref{#1})}
\def\Eq#1{Eq.~(\ref{#1})}

\def\eqs#1#2{eqs.~(\ref{#1}) and (\ref{#2})}
\def\eqss#1#2#3{eqs.~(\ref{#1}), (\ref{#2}) and (\ref{#3})}

\def\Ref#1{ref.~\cite{#1}}

\def\Rrefs#1{Refs.~\cite{#1}}
\def\Refs#1{refs.~\cite{#1}}
\newcommand{\bra}[1]{\langle #1|}
\newcommand{\ket}[1]{|#1\rangle}
\newcommand{\braket}[2]{\langle #1|#2\rangle}
\title{Anatomy of the Higgs boson decay into two photons in the unitary gauge}

\author{Athanasios Dedes\footnote{email: {\tt adedes@cc.uoi.gr}}~  and 
Kristaq Suxho\footnote{email: {\tt csoutzio@cc.uoi.gr}}}
\affil{Division of Theoretical Physics, 
 University of Ioannina,  GR 45110, Greece}
\date{\today}                                           % Activate to display a given date or no date
\begin{document}

\maketitle

\begin{abstract}
In this work, we review and clarify computational issues about the $W$-gauge boson 
one-loop contribution
to the $H\to \gamma\gamma$ decay amplitude, in the unitary gauge and in the Standard Model. 
We find that highly divergent 
integrals depend upon the choice of shifting  momenta  with arbitrary vectors.
One particular combination of these arbitrary vectors reduces the superficial divergency 
%by  at most six powers.
down to a logarithmic one.
The remaining  ambiguity is then fixed by exploiting gauge invariance  and the Goldstone Boson Equivalence Theorem. Our method is strictly realised in four-dimensions. The result for the 
amplitude agrees with the ``famous'' one obtained using dimensional regularisation (DR) in
 the limit  $d\to 4$, 
where $d$ is the number of spatial dimensions
in Euclidean space. 
At the exact equality $d=4$,   a three-sphere 
surface term appears that renders the Ward Identities and the equivalence theorem 
inconsistent.  We also examined a recently proposed four-dimensional regularisation
scheme and found agreement with the DR outcome.
\end{abstract}

\newpage
%%%%%%%%%%%%%%%%%%%%%%%%%%%%%%%%%
\section{Introduction}
\label{intro}

Today one of the main focal points at the Large Hadron Collider (LHC) 
is to search for the Higgs boson ($H$)~\cite{Higgs:1964pj,Englert:1964et,Guralnik:1964eu}
through its decay into two photons, $H\to \gamma\gamma$ 
(for reviews see~\cite{Gunion:1989we,Djouadi:2005gi}). 
Indeed, the recent~\cite{ATLAS:2012gk,CMS:2012gu}  observation by ATLAS and CMS 
experiments of a resonance,  that could be the Standard Model (SM) Higgs
boson, is based on data mainly driven  by $H\to \gamma\gamma$.
In the  (SM)\cite{Weinberg:1967tq,Glashow,Salam}, 
this particular decay process
goes  through loop induced diagrams involving either charged fermions or $W$-gauge bosons. 
Their calculation was first performed in \Ref{Ellis:1975ap} in the limit of light Higgs mass
$m_{H}\ll m_{W}$,
using dimensional regularisation in the 't Hooft-Feynman gauge. Since then, there are 
numerous works spent on improving this calculation including finite Higgs mass 
effects in linear and non-linear gauges~\cite{Ioffe:1976sd,Shifman:1979eb,Gavela:1981ri}, 
different regularisation schemes~\cite{Huang:2011yf,Shao:2011wx,Bursa:2011aa,Piccinini:2011az} 
and/or different gauge choices~\cite{Marciano:2011gm}.
%Why there should be one more calculation on  $H\to \gamma\gamma$ amplitude?

%%%%%%%%% MOTIVATION %%%%%%%%%%%%%
%%%%%%%%%%%%%%%%%%%%%%%%%%
\begin{figure}[t] %  figure placement: here, top, bottom, or page
   \centering
   \includegraphics[width=4in]{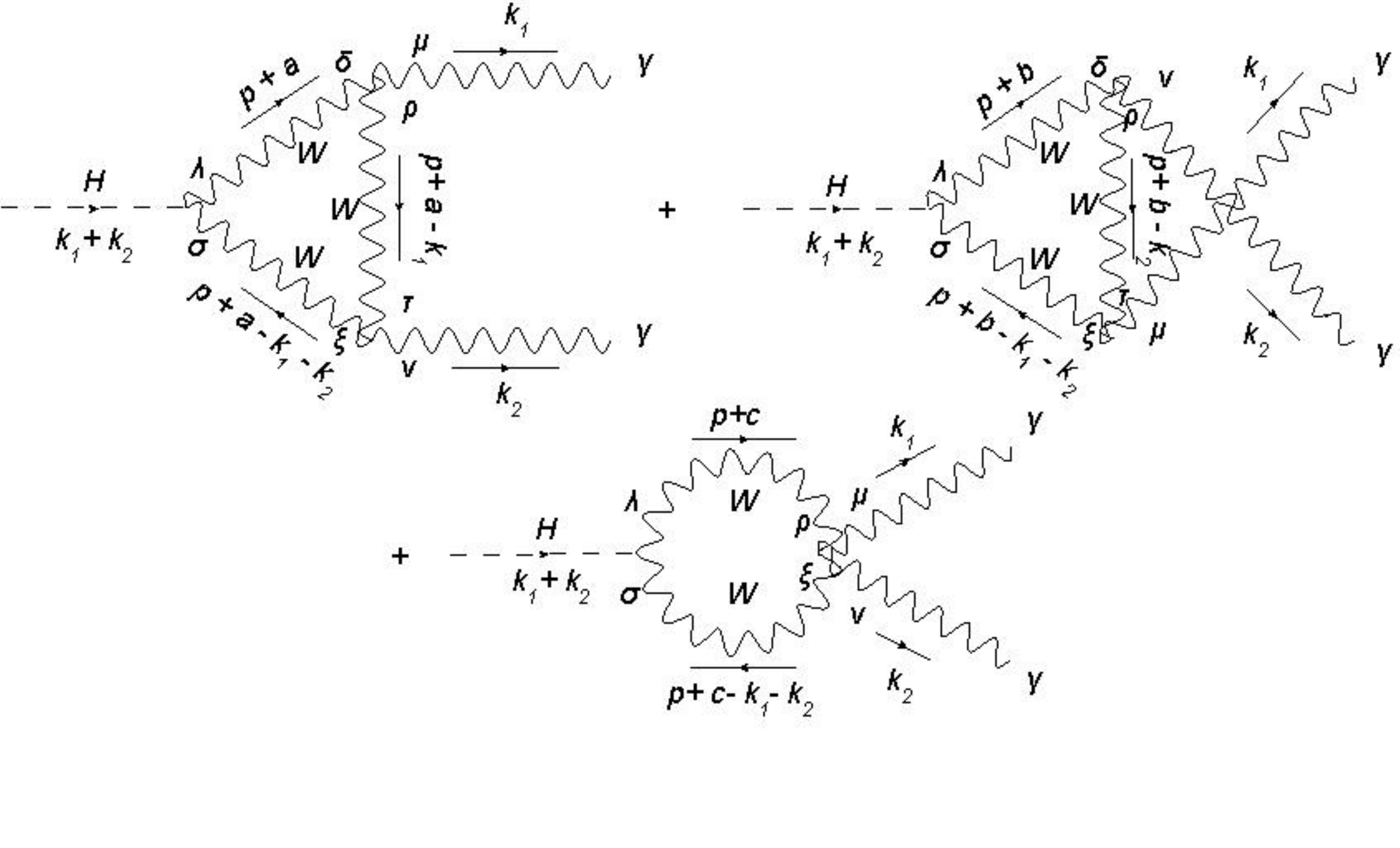}
   %5\includegraphics[width=4in]{diagrams1.jpg} 
   \caption{$W$-gauge boson contribution to the $H \to \gamma \gamma$ amplitude. Momentum
   flow together with relevant shift vectors are indicated.}
   \label{fig1}
\end{figure}
%%%%%%%%%%%%%%%%%%%%%%%%%%%%

The $H\to \gamma\gamma$ amplitude is originated, in broken (unbroken) phase, 
by a dimension-5 (dimension-6) SM gauge invariant
operator(s) and, therefore, its expression, within a renormalizable theory, must be finite, gauge
invariant and  independent of any gauge choice. The amplitude should also be consistent
with the Goldstone Boson Equivalence 
Theorem (GBET)~\cite{Cornwall,Vayonakis,Lee} since the SM is a 
renormalizable, spontaneously broken, gauge  field theory. 

A problem arises when the $W$-gauge boson  contribution 
(see Fig.~\ref{fig1})
to $H\to \gamma\gamma$  produces ``infinite'' results at intermediate steps. These 
problems are usually treated
by using a gauge invariant regulator method, e.g., dimensional regularization. 
In the unitary gauge~\cite{Weinberg:1973ew},
this indeterminacy is more pronounced and more difficult to handle with\footnote{However,
using DR and unitary gauge with modern computer algorithms this may not be a hard
problem today~\cite{Marciano:2011gm}.}
due to the  particular form of the $W$-gauge boson propagator. 
On the other hand it is much simpler to work with only few diagrams, that involve
physical particle masses, rather than
many.

More specifically, in the unitary gauge, one encounters 
divergencies up to the sixth power. It is well known that,  in four-dimensions,
shifting momenta in 
integrals  that are more than logarithmically divergent is a ``tricky business''
- recall the calculation of linearly divergent fermion triangles in chiral anomalies~\cite{Adler,Bell} -
that requires keeping track of several ``surface'' terms for these integrals. 
%Therefore,
%at first, dimensional regularisation technics that involve ``shifting momentum'' procedures
%through Feynman parameters are not well  
%suited \marginpar{this is wrong!}
%in calculating the $H\to \gamma\gamma$ amplitude in unitary gauge. 
There  is also the situation we face here where  
apparent logarithmically divergent integrals  turn out to be finite 
but discontinuous at $d=4$.
%
%Second, is a case we encounter in this article: a logarithmically divergent integral
%which is discontinuous at exactly $d=4$. 

%%%%%%%%% CALCULATIONAL STEPS %%%%%%%
We would like to bypass those ambiguities and at the same time to present a 
``regularisation'' method, 
by performing the calculation for
the $H\to \gamma\gamma$ amplitude strictly in 4-dimensions and in the physical 
unitary  gauge. Our method is similar to the one used elsewhere 
for calculating triple gauge
boson amplitudes~\cite{Rosenberg,Dedes:2012me}, 
or Lorentz non-invariant amplitudes~\cite{Jackiw:1999qq},
and consists of three steps:
\begin{enumerate}
\item We write down the most general Lorentz invariant $H\to \gamma\gamma$ amplitude.
\item We introduce  arbitrary vectors that account for the ``shifting momentum'' indeterminacy.
 We show that a particular choice of those ``shifting vectors'' 
 cancel higher powers of infinities leaving still behind
 at most logarithmically divergent integrals that are  treated as undetermined variables.
\item We exploit physics, i.e., gauge invariance (Ward Identities) and the GBET 
in order to fix the last undetermined  variables.
\end{enumerate}
%%%%%%%%%%%%%%%%%%%%%
This method is quite general within a renormalizable theory
and can be applied to other observables too. 
Following these steps we arrive at the same result for the $H\to \gamma\gamma$ amplitude
obtained by J.~Ellis et.al~\cite{Ellis:1975ap} and by M.~Shifman et.al~\cite{Shifman:1979eb}
almost 35 years ago. Our analysis, among other issues,  highlights that the recent 
observation~\cite{ATLAS:2012gk,CMS:2012gu} of the
$H\to \gamma\gamma$  at the LHC 
%in (almost) agreement with the above physics setup 
signifies the validity of the Goldstone Boson Equivalence  Theorem. 
As a further clarification we also make a  remark on the direct calculation in the following three
cases: we first perform the integrals in exactly $d=4$ 
(with no regularisation method beyond the one discussed in point 2 above),
second, by exploiting  Dimensional Regularisation (DR) as defined in 
\Refs{'tHooft:1972fi,Collins}
and then taking the limit $d\to 4$, and finally third by using a four-dimensional
regularization scheme  introduced in \Ref{Pittau:2012zd}.

Our calculation is complementary to, but somewhat 
different than, the two existing ones~\cite{Gastmans:2011ks,Gastmans:2011wh,Marciano:2011gm}
performed  in the unitary gauge.  
It is not our intend to redo the calculation 
in unitary gauge with DR as in \Ref{Marciano:2011gm}. On the contrary,
we want to clarify subtle issues related to the amplitude 
in unitary gauge and $d=4$ raised in part by \Refs{Gastmans:2011ks,Gastmans:2011wh}. 
We find, using arbitrary vectors, 
that divergencies (up to 6th power) are reduced  down to 
logarithmic ones. This is a new result that is not obvious
when working in unitary gauge and cannot be seen 
when using dimensional regularisation. 
This fact  was stated incorrectly in \Refs{Gastmans:2011ks,Gastmans:2011wh}.

%This is the third calculation of the $H\to \gamma\gamma$ one-loop amplitude
%performed in unitary gauge. First, very recently \Ref{Gastmans}
%directly in four-dimensions and the other in \Ref{Marciano} using DR. 
%Our calculation, though different, agrees with the result of the later paper.  

The outline of the paper is as following:
in section~\ref{calc} we present the calculation 
of the $W$-loop contribution\footnote{Note that the 
calculation of the fermion  triangle contribution is well
defined i.e., it is independent of arbitrary vectors and finite. We are not
going to repeat this calculation here and  refer the reader to the reviews 
in \Refs{Gunion:1989we,Djouadi:2005gi}.}
to $H\to \gamma\gamma$ amplitude, its ambiguities and the resolution
within  physics arising from GBET. 
Next, in section~\ref{sec:fdr} we examine details of the amplitude calculation 
within an alternative and recently 
proposed four dimensional regularization scheme~\cite{Pittau:2012zd},
the one that resembles most closely the symmetry  approach taken here. 
In section~\ref{discussion} we discuss other possible 
physical setups plus experiment that may help to resolve inconsistencies. 
We conclude in section~\ref{conclusions}. There are two Appendices : A that contains 
some intermediate  formulae  and B where we present the details about  the origin of 
surface terms in four dimensions.

%%%%%%%%%%%%%%%%%%%%%%%%%%%%%%%%
\section{The $W$-loop contribution to $H\to \gamma \gamma$ in SM}
\label{calc}

The most general, Lorentz and CP- invariant, form of the of-shell $H\to \gamma \gamma$ amplitude
is,
%%%%%%%%%%%
\begin{equation}
%\mathcal{M}^{\mu\nu}(k_{1},k_{2}) \ = \ 
\mathcal{M}_{1}\, g^{\mu\nu} + 
\mathcal{M}_{2}\, k_{1}^{\nu}\, k_{2}^{\mu} +
\mathcal{M}_{3}\, k_{1}^{\mu}\, k_{2}^{\nu} + \mathcal{M}_{4}\, k_{1}^{\mu}\, k_{1}^{\nu} +
\mathcal{M}_{5}\, k_{2}^{\mu}\, k_{2}^{\nu} \;, \label{gam1}
\end{equation}
%%%%%%%%%%%%  
where $k_{1}$ and $k_{2}$ are the outgoing photon momenta shown in Fig.~\ref{fig1}, and 
the coefficients $\mathcal{M}_{i=1..5}\equiv \mathcal{M}_{i=1..5}(k_{1},k_{2})$ 
are scalar functions of 
$k_{1}^{2}, k_{2}^{2}$, and $k_{1}\cdot k_{2}$. By considering that all particles are on-mass-shell,
that is $k_{1}^{2} = k_{2}^{2}=0\;, k_{1}\cdot k_{2} = m_{H}^{2}/2\;, 
k_{1}\cdot \epsilon^{*}(k_{1})=0\;, k_{2}\cdot \epsilon^{*}(k_{2})=0$,
%k_{1}^{\mu}\epsilon_{\mu}^{*}(k_{1}) =0\;,
%k_{2}^{\nu}\epsilon_{\nu}^{*}(k_{2}) =0$, 
we obtain an amplitude 
$\mathcal{M} = \mathcal{M}^{\mu\nu} \epsilon_{\mu}^{*}(k_{1}) \epsilon_{\nu}^{*}(k_{2})$
with only two, undetermined (for the time being),  coefficients,
%%%%%%%%%%%%
\begin{equation}
\mathcal{M}^{\mu\nu}  =
 \mathcal{M}_{1}\, g^{\mu\nu} \ + \ \mathcal{M}_{2}\, k_{1}^{\nu} \, k_{2}^{\mu}  \;\;. \label{gam2}
\end{equation}
%%%%%%%%%%%%%%%%%%%%
In unitary gauge, the Feynman diagrams that contribute to 
$\mathcal{M}_{1}$ and $\mathcal{M}_{2}$  are displayed in 
Fig.~\ref{fig1}. 
In order to calculate them, 
we introduce three arbitrary four-vectors $a,b$ and $c$, one for each diagram. 
These vectors
shift the integration  momentum, i.e., $p\rightarrow p + a$ for the first diagram, 
$p\rightarrow p + b$ for the second diagram and $p\rightarrow p+c$ for the third diagram.
As we shall see, these arbitrary vectors 
operate as regulators capable to handle highly divergent integrals related to unitary gauge choice. 
Furthermore, the vectors $a,b$ and $c$, linearly depend upon the external momenta $k_{1}$ 
and $k_{2}$. Hence $a,b$ and $c$ are not linearly independent [{\it c.f.} \eq{con1}]. 
This  is an important fact leading to the cancellation of infinities.

We first calculate the less divergent part of $\mathcal{M}^{\mu\nu}$ in \eq{gam2}
which  is the $\mathcal{M}_{2}$ coefficient\footnote{The coefficient
$\mathcal{M}_{1}$ will be fixed later on by 
the requirement of gauge invariance.}. 
By naive power counting, we see that $\mathcal{M}_{2}$ diverges 
 by at most four powers.
Then we perform the Feynman integral calculations strictly in 4-dimensions. 
For reasons that will become clear later, we shall keep  the number
of dimensions general in all intermediate steps of the calculation i.e., 
$g^{\mu\nu}g_{\mu\nu}=d$. As we will see, $d$ contributes only
in finite pieces of $\mathcal{M}_{2}$ [{\it c.f.} \eq{ma2}]\footnote{On the contrary, 
we shall see that 
there are non-trivial $d$-contributions into $\mathcal{M}_{1}$-coefficient.}. 

With all the above definitions, we can 
write down the total amplitude in the form
%%%%%%%%%%%%%%%%%%%%%%%%%%%%
\begin{eqnarray}
\mathcal{M}^{\mu\nu}  &\sim&  \int \frac{d^{4}p}{(2\pi)^{4}}\: \bigl [
 \mathcal{A}_{11} \, g^{\mu\nu} \nonumber \\
&+& \mathcal{A}_{21}\, (p+a)^{\mu}\, (p+a)^{\nu}    +
\mathcal{A}_{22} \, (p+b)^{\mu}\, (p+b)^{\nu} +
\mathcal{A}_{23} \, (p+c)^{\mu}\, (p+c)^{\nu} \nonumber \\
&+& \mathcal{A}_{31} \, (p+a)^{\mu} \, k_{1}^{\nu} +
\mathcal{A}_{32} \, (p+b)^{\mu} \, k_{1}^{\nu} + \mathcal{A}_{33} \, (p+c)^{\mu} \, k_{1}^{\nu}
\nonumber \\
&+& \mathcal{A}_{41} \, (p+a)^{\nu} \, k_{2}^{\mu} + 
\mathcal{A}_{42} \, (p+b)^{\nu} \, k_{2}^{\mu} + \mathcal{A}_{43} \, (p+c)^{\nu} \, k_{2}^{\mu}
\nonumber \\
&+&\mathcal{A}_{51} \, k_{2}^{\mu} \, k_{1}^{\nu}\, \bigr ]\;, \label{mat}
\end{eqnarray}
%%%%%%%%%%%%%%%%%%%%%%%%%
where the coefficients 
$\mathcal{A}_{ij} = \mathcal{A}_{ij}(p^{n};k_{1},k_{2};a;b;c)$ with $-6\le n\le 0$,
 are given in \ref{sec:appA},
and the $\sim$ sign is the proportionality factor: $-\frac{2i  e^{2}}{v}$.
Note that $\mathcal{M}^{\mu\nu}$ is a (superficially) 
6th power divergent  amplitude  in the unitary gauge. 
$\mathcal{A}_{11}$ in \eq{mat} solely contributes to $\mathcal{M}_{1}$ while all other 
$\mathcal{A}$-elements contribute
to both $\mathcal{M}_{1}$ and/or $\mathcal{M}_{2}$ in \eq{gam2}. 

First we focus on the calculation of the ``less divergent'' coefficient
$\mathcal{M}_{2}$ of \eq{gam2}.
Based on naive power counting, we observe that the
$\mathcal{A}_{21},\mathcal{A}_{22},\mathcal{A}_{23}$-terms in \eq{mat},
 lead to at the most quartic divergent integrals.
However, when adding all these pieces together, we find that quartic divergent integrals vanish 
for every arbitrary vectors $a,b$ and $c$ leaving behind 
an expression with  integrals of third power (in momenta) plus integrals with smaller 
divergencies.
Then the cubically divergent integrals are proportional to all possible Lorentz
invariant combinations like: $[(a+b-2 c)\cdot p]\, p^{\mu} p^{\nu}$, 
$[(a+b-2 c)^{\nu} p^{\mu}]\, p^{2}$
and $[(a+b-2 c)^{\mu} p^{\nu}]\, p^{2}$. 
Therefore, choosing 
%%%%%%%%%%%%%%%%%%
\begin{equation}
a+ b -2\, c \ = \ 0 \;,\label{con1}
\end{equation}
%%%%%%%%%%%%%%%%%%%
we ensure that third order divergent integrals related to $\mathcal{A}_{21},\mathcal{A}_{22},\mathcal{A}_{23}$-terms, vanish identically. In the same way,  by naive power counting,
$\mathcal{A}_{31}$ and  $\mathcal{A}_{33}$-terms - these terms in \eq{mat}
together with $\mathcal{A}_{32}$ contribute solely to $\mathcal{M}_{2}$ -
lead again to at most third order divergent integrals.
However,  in the sum of $\mathcal{A}_{31}$ and $\mathcal{A}_{33}$-terms in \eq{mat}, 
third order divergent integrals 
vanish for arbitrary $a,b$ and $c$, leading to an expression, 
that when added to $\mathcal{A}_{32}$-term,
consists of at most  quadratically divergent integrals, proportional to 
$[(c-a)\cdot p]\, p^{\mu} k_{1}^{\nu}$ and $[(c-a)^{\mu}k_{1}^{\nu}]\, p^{2}$. We choose,
%%%%%%%%%%%%%%%%%%
\begin{equation}
c-a = \ 0 \;,\label{con2}
\end{equation}
%%%%%%%%%%%%%%%%%%%
for the quadratically divergent integrals to vanish. Likewise, when we add 
$\mathcal{A}_{42}$ and  $\mathcal{A}_{43}$-terms - these terms, together 
with $\mathcal{A}_{41}$,
solely contribute to $\mathcal{M}_{2}$
in \eq{gam2} - the third order divergent integrals vanish  for every choice of $a,b,c$ 
leading to an expression, that when added to $\mathcal{A}_{41}$,
consists of at most quadratically divergent integrals proportional to 
$[(c-b)\cdot p]\, p^{\nu} k_{2}^{\mu}$ and $[(c-b)^{\nu} k_{2}^{\mu}]\, p^{2}$. 
Therefore,
we choose
%%%%%%%%%%%%%%%%%%
\begin{equation}
c-b = \ 0 \;,\label{con3}
\end{equation}
%%%%%%%%%%%%%%%%%%%
for infinities 
%arising from these quadratically divergent integrals 
to vanish identically. From \eqss{con1}{con2}{con3} we arrive at the
final relation among the three introduced vectors:
%%%%%%%%%%%%%%%%%%
\begin{equation}
a \ = \ b \ = \ c \;.\label{con}
\end{equation}
%%%%%%%%%%%%%%%%%%%
\Eq{con} suggests
 that the rest of the divergent integrals depend by, at most, one arbitrary vector, say the $a$-vector.
Note that $\mathcal{A}_{51}$ contributes only to the finite part of $\mathcal{M}_{2}$.
Now, if we impose conditions (\ref{con}) onto  the remaining expressions for 
$\mathcal{A}_{21},\mathcal{A}_{22},...,\mathcal{A}_{51}$-terms of \eq{mat}, 
we find that \emph{all} quadratically and linearly divergent integrals
vanish, independently of the direction of the $a$-vector. 
We stress here the fact that \emph{the cancellation of divergencies down to logarithmic ones
is a highly non-trivial, almost ``miraculous'',  result. These cancellations only
take place for a particular choice of the momentum-variable shift vectors, 
[\eq{con}]}\footnote{As a corollary, if for instance,
we had split the $WW\gamma$-vertex into three pieces, each one associated
with three different arbitrary vectors, then  the generalised condition (\ref{con})
  would again downgrade
the divergency of the amplitude to a logarithmic one.}.  
Of course this is an expected outcome for an observable in a renormalizable 
theory.

Our final result contains \emph{at most logarithmically divergent integrals}.
\footnote{This result is different with the one obtained in 
 \Refs{Gastmans:2011ks,Gastmans:2011wh}, where there are  remaining quadratically and linearly
 divergent terms.
 Following eq.(11) in Gastmans {\it et.al} paper~\cite{Gastmans:2011ks}
or eq.(3.36) in their sequel paper~\cite{Gastmans:2011wh},
and unless there is a typo in both their formulae, we find
that there is a missing quadratically divergent term proportional to $k^2 k_\mu k_\nu$.
 Even in the case this is a typo,  their formulae
contain a linearly divergent term that is also referred to by the authors
 claiming that this  last term reduces to a logarithmically divergent integral
by changing the internal momentum $k \to -k$ and further manipulating  the integral. 
Note here that in our calculation all divergent integrals are reduced 
to at most logarithmically divergent ones \emph{without} any further manipulation nor
any assumption  other than \eq{con} for every arbitrary vectors $a,b$, and $c$.
Finally, the authors in \Rrefs{Gastmans:2011ks,Gastmans:2011wh} 
have made a specific choice for routing the internal loop momentum
corresponding to $a=b=c=\frac{1}{2}(k1+k2)$ and therefore their result should be the same
with ours.}
Despite of the fact that the resulting expressions so far contain the shift $p+a$ instead of $p$
with an arbitrary vector $a$, 
its presence is irrelevant since logarithmically divergent integrals are momentum-variable
 shift independent~\cite{Rohrlich}. 
Summing up  all the above contributions to $\mathcal{M}_{2}$, 
we find a particularly nice and symmetric form
 for $\mathcal{M}^{\mu\nu}$, 
%%%%%%%%%%%%%%%%%%
\begin{eqnarray}
\mathcal{M}^{\mu\nu} 
&\sim & \int \frac{d^{4}p}{(2\pi)^{4}}\, p^{\mu} p^{\nu}\, \biggl \{
\frac{4(d-1)\, m^{2}_{W}+
 2 \, m^{2}_{H}}{[p^{2}-m^{2}_{W}][(p-k_{1})^{2}-m^{2}_{W}][(p-k_{1}-k_{2})^{2}-m^{2}_{W}]} \nonumber \\[3mm]
 &+& 
 \frac{4(d-1)\ m^{2}_{W}+
 2 \, m^{2}_{H}}{[p^{2}-m^{2}_{W}][(p-k_{2})^{2}-m^{2}_{W}][(p-k_{1}-k_{2})^{2}-m^{2}_{W}]}
 \bigg \} \nonumber \\[3mm]
 %%%%%%%%%%%%%%%%%%%%%%%%%
 &+& \int \frac{d^{4}p}{(2 \pi)^{4}}\,p^{\mu}
 k^{\nu}_{1}\,\bigg \{\frac{-4(d-1)\, m^{2}_{W}-
 4 \, (p\cdot k_{2})}{[p^{2}-m^{2}_{W}][(p-k_{1})^{2}-m^{2}_{W}][(p-k_{1}-k_{2})^{2}-m^{2}_{W}]}
  \nonumber \\[3mm]
 &+& \frac{-4 \, (p\cdot k_{2})}{[p^{2}-m^{2}_{W}][(p-k_{2})^{2}-m^{2}_{W}][(p-k_{1}-k_{2})^{2}-m^{2}_{W}]}\bigg \}\nonumber \\[3mm]
%%%%%%%%%%%%%%%%%%%%%%
&+& \int \frac{d^{4}p}{(2 \pi)^{4}}\, p^{\nu}
 k^{\mu}_{2}\,\bigg \{ \frac{-
 4 \, (p\cdot k_{1})}{[p^{2}-m^{2}_{W}][(p-k_{1})^{2}-m^{2}_{W}][(p-k_{1}-k_{2})^{2}-m^{2}_{W}]}
  \nonumber \\[3mm]
 &+& \frac{-4(d-1)\, m^{2}_{W}-4 \, (p\cdot k_{1})}{[p^{2}-m^{2}_{W}][(p-k_{2})^{2}-m^{2}_{W}][(p-k_{1}-k_{2})^{2}-m^{2}_{W}]}\bigg \}\nonumber \\[3mm] 
 %%%%%%%%%%%%%%%
 &+&\int \frac{d^{4}p}{(2 \pi)^{4}}\, k^{\nu}_{1}
 k^{\mu}_{2}\, \bigg \{\frac{6\, m^{2}_{W}+
 2 \, p^{2}}{[p^{2}-m^{2}_{W}][(p-k_{1})^{2}-m^{2}_{W}][(p-k_{1}-k_{2})^{2}-m^{2}_{W}]}
  \nonumber \\[3mm]
 &+&\frac{6 \, m^{2}_{W}+2 \, p^{2}}{[p^{2}-m^{2}_{W}][(p-k_{2})^{2}-m^{2}_{W}][(p-k_{1}-k_{2})^{2}-m^{2}_{W}]}\bigg \}\;.
 \label{M1}
 %%%%%%%%%%%%%%%%%%%%%%%%%%%%
\end{eqnarray}
%%%%%%%%%%%%%%%%% %
Introducing Feynman parameters, shifting momentum variable from $p$ to $\ell$ 
 and ignoring all 
 terms\footnote{These terms will be used later in arriving at  \eq{m1}.} that contribute to 
$\mathcal{M}_{1}$  we find
that the contribution to $\mathcal{M}_{2}$ in \eq{gam2} arises solely from the term,
%%%%%%%%%%%%%%%%%%%%%%%
\begin{eqnarray}
\mathcal{M}_{2} \, k_{1}^{\nu}\, k_{2}^{\mu}  &\sim & 
 8 \,
\int_{0}^{1} dx  \int_{0}^{1-x} dy  \int \frac{d^{4}\ell}{(2\pi)^{4}} \:
\frac{\ell^{2}\, k_{1}^{\nu}\, k_{2}^{\mu}  - 
2 \,(\ell \cdot k_{2})\, \ell^{\mu} \, k_{1}^{\nu} - 
2  \, (\ell \cdot k_{1}) \,
\ell^{\nu} \, k_{2}^{\mu}}{(\ell^{2} -\Delta)^{3}} \nonumber \\[5mm]
&+&
8\: m_{W}^{2}\, \int_{0}^{1} dx  \int_{0}^{1-x} dy  \int \frac{d^{4}\ell}{(2\pi)^{4}} \:
\frac{3-2\,(d-1)\, x\, (1-x-y)}{(\ell^{2}-\Delta)^{3}}\, k_{1}^{\nu} \, k_{2}^{\mu} \;,
\label{ma2}
%%%%%%%%%%%%%%%%
\end{eqnarray}
%%%%%%%%%%%%%%%%%
with $\Delta = x (x+y-1)\, m_{H}^{2} + m_{W}^{2}$. 
Obviously, the first integral in \eq{ma2} is (superficially) logarithmically
divergent while the second one is finite.
The number of dimensions $(d)$ appears
only at the finite integral and therefore we can fearlessly set $d=4$ everywhere.
This means that we do not use dimensional regularisation in what follows (see however
the  discussion below).
We state here few additional remarks to be exploited later on: 
a) we observe that 
the top line in the integrand of \eq{ma2}
\emph{does not vanish} in the limit $m_{W}^{2}\to 0$  and, b) despite of appearances 
in \eq{M1}, there is no $m_{H}^{2}$ in the numerators  of the subsequent expression
\eq{ma2}. The whole $m_{H}^{2}$ contribution arises from the denominator's $\Delta$-term.

Our next step is to parametrize the logarithmically divergent 
integral in \eq{ma2} by an unknown, dimensionless, parameter $\lambda$ 
to be determined later by a physical argument. So we define,
%%%%%%%%%%%%%%%%%%
\begin{eqnarray}
\int_{0}^{1} dx  \int_{0}^{1-x} dy  \int \frac{d^{4}\ell}{(2\pi)^{4}} \:
\frac{\ell^{2}\, k_{1}^{\nu}\, k_{2}^{\mu}  - 
2 \,(\ell \cdot k_{2})\, \ell^{\mu} \, k_{1}^{\nu} - 
2  \, (\ell \cdot k_{1}) \,
\ell^{\nu} \, k_{2}^{\mu}}{(\ell^{2} -\Delta)^{3}} 
%\nonumber \\[3mm]
 \equiv \frac{i \lambda}{4(4\pi)^{2}}\,
k_{1}^{\nu}\, k_{2}^{\mu} \;. \nonumber \\ \label{lam}
\end{eqnarray}
%%%%%%%%%%%%%%%%%%%%%%%%
An important parenthesis here. We 
could of course promote $d^{4}\ell \to d^{d}\ell$ and use dimensional 
regularisation~\cite{'tHooft:1972fi} by exploiting  symmetric integration
$\ell^{\mu}\ell^{\nu} \to \frac{1}{d}\ell^{2} g^{\mu\nu}$ in $d$-dimensions. 
In this case, and after taking the limit $d\to 4$,
one finds $\lambda = -1$ which  is finite and non-zero, 
and, agrees with the one we find below in \eq{216} after imposing 
the GBET condition. 
This is also the result found in the original~\Refs{Ellis:1975ap,Ioffe:1976sd,Shifman:1979eb}. 
However, according to~\Refs{Gastmans:2011ks,Gastmans:2011wh}, 
the integral in \eq{lam} is discontinuous at $d=4$; in fact, when symmetric integration, $\ell^{\mu}\ell^{\nu} \to \frac{1}{4}\ell^{2} g^{\mu\nu}$
 in $d=4$ is used,  one finds instead $\lambda=0$.
This is also understood in a slightly different context. It has long been 
known~\cite{Rohrlich,Pugh:1969kn,Elias:1982sq} that shifts of integration variables
in linearly (and above) divergent integrals  
are accompanied by ``surface'' terms that appear only in four dimensions -- 
a famous example being the integrals in chiral anomaly triangle graphs.
For our purpose here lets start with the following shift of variables in a linearly 
divergent integral that has
been generalised~\cite{Elias:1982sq} to work in $2\omega$-dimensions following 
the expression,
%%%%%%%%%%%%%%%%%%%%
\begin{equation}
\int d^{2\omega}\ell \; \frac{\ell_{\mu}}{[(\ell - k)^{2} - \Delta]^{2}} -  
\int d^{2\omega}\ell \; \frac{(\ell + k)_{\mu}}{(\ell^{2} - \Delta)^{2}} \ = \
- \frac{i\pi^{2}}{2}\, k_{\mu}\, \delta_{\omega,2} \;,\label{mann}
\end{equation}  
%%%%%%%%%%%%%%%%%%%%%%%
 that is valid for $\omega<5/2$ and $\Delta$  constant, possibly dependent on
 Feynman parameters, like the one given below \eq{ma2}, and $k_{\mu}$ is an arbitrary constant
  four vector. By taking the derivative, 
 $\frac{\partial}{\partial k^{\nu}}$, of both sides in \eq{mann} and shifting the integration
 variable for the logarithmically divergent integral encountered, and evaluating
 the finite one, we easily arrive at
 %%%%%%%%%%%%%%%
 \begin{equation}
 \int d^{2\omega}\ell \; \frac{\ell^{2}\, g_{\mu\nu} - 
 4\,\ell_{\mu}\, \ell_{\nu}}{(\ell^{2} - \Delta)^{3}} \ = \
 -\frac{i\pi^{2}}{2}\, g_{\mu\nu} \; \biggl (\frac{\pi^{\omega-2}\, \Gamma(3-\omega)}{\Delta^{2-\omega}}- \delta_{\omega,2}\biggr )  \;.    
 \label{mann2}
 \end{equation}
 %%%%%%%%%%%%%%%%%%
 For an alternative and detailed proof of \eq{mann2}, see \ref{sec:appB}.\footnote{The same
 result is obtained by standard algebraic tricks. We would like to thank R. Jackiw 
 for communicating his calculation to us.} 
 Applying  \eq{mann2} to $\ell^{\sigma}\ell^{\rho}$ terms of \eq{lam} with 
 $\frac{d^{4}\ell}{(2\pi)^{4}}\to \frac{d^{2\omega}\ell}{(2\pi)^{2\omega}}$,  we find, 
 %%%%%%%%%%%%%%%
 \begin{eqnarray}
\lambda = \biggl \{ \begin{array}{c} \hspace*{-1.6cm} 0 \;, \qquad \omega = 2 \\[3mm]
-1 \,, \qquad \omega = 2 - \epsilon \quad \mathrm{(DR)}\end{array} \;. 
\label{lamdef}
\end{eqnarray}
 %%%%%%%%%%%%%%%%%%%%
This is consistent with  the symmetric integration in 4-dimensions ($\omega=2$), {\it but},
is also consistent  with the usual tabulated textbook result~\cite{Peskin} from 
dimensional regularisation in $4-2\epsilon$-dimensions ($\omega = 2-\epsilon$).
\Eq{lamdef} shows that $\lambda$ is discontinuous at $d=2\omega=4$.
Then the Question arises: {\it which $\lambda$ to believe in?} Answer: {\it the one
that is indicated by well defined, calculable, boundary conditions and symmetries of the underlying theory.}
 
%the 
%exact  $d=4$ choice seems to be problematic due to .....  resulting in $\lambda =0$. 
The above parenthesis to our 
calculation motivates us to avoid the direct calculation of integral (\ref{lam})
but set $d=4$ everywhere and 
treat $\lambda$ as an unknown parameter to be defined later 
within a physical context or experiment.
%
%There is however a question  whether the exact $d=4$ choice is
%permissible or not. 
%For a discussion at this point see \Ref{Gastmans:2011ks,Gastmans:2011wh}. 
%We want to avoid this question  and that is why we want to determine the result 
%within a physical context [see below].
%
%
Substituting \eq{lam} into \eq{ma2} we  arrive at 
%%%%%%%%%%%%%%%
\begin{equation}
\mathcal{M}_{{2}}  \sim \frac{i}{8\pi^{2}}\, \biggl \{
\lambda -  6\: m_{W}^{2}\, \int_{0}^{1} dx  \int_{0}^{1-x} dy   \:
\frac{1-2 x (1-x-y)}{\Delta} \biggr \}  \;.\label{ma22}
\end{equation}
%%%%%%%%%%%%%%%%%%%%%
Evaluating the double finite integral in \eq{ma22}, and restoring the proportionality
factor   given below \eq{mat}, we obtain,
%%%%%%%%%%%%%%%%%%%%%%
\begin{equation}
\mathcal{M}_{{2}} = -\frac{e^{2} g}{(4\pi)^{2} m_{W}} \:
\biggl \{ -2 \lambda + \biggl [ 3\: \beta + 3 \: \beta\: (2-\beta)\: f(\beta) \biggr ] \biggr \}  \;, \label{eq:2.12}
\end{equation}
%%%%%%%%%%%%%%%%%%%
where 
%%%%%%%%%%%%%%%%%%%%%%%
\begin{equation}
\beta = \frac{4 m_{W}^{2}}{m_{H}^{2}}\;, \quad {\rm and }, \quad 
f(\beta) = \Biggl \{ \begin{array}{c} \arctan^{2} \left (\frac{1}{\sqrt{\beta -1}} \right )\;, \;
\beta \ge 1  \\[3mm] 
-\frac{1}{4} \: \left [\ln \left (\frac{1+\sqrt{1-\beta}}{1-\sqrt{1-\beta}}\right ) -i \: \pi \right ]^{2}\;,\;
\beta <1 
\end{array} \;. \label{bfb}
\end{equation}
%%%%%%%%%%%%%%%%%

%\subsection{Goldstone Boson Equivalence Theorem (GBET)}

Our final step is to determine the unknown parameter $\lambda$ in \eq{eq:2.12}.
For this we need physics  that reproduces $\mathcal{M}_{{2}}$ 
in a different and unambiguous way.
One choice, probably not the only one, is to adopt  the  
Goldstone Boson Equivalence Theorem (GBET)~\cite{Cornwall,Vayonakis,Lee} which states that
{the amplitude for emission 
or absorption of a longitudinally polarised $W$ at high energy becomes equivalent 
to the emission or absorption of the Goldstone boson that was eaten}. 
Mathematically, this is written by an equation~\cite{Bagger:1989fc},
\begin{equation}
S[\, W_{L}^{\pm}, \; \mathrm{physical}\, ] = i^{n} \, S[\, s^{\pm},\; \mathrm{physical}\,]\;,
\label{GBET}
\end{equation}
which says that the S-matrix elements for the scattering 
of the \emph{physical} longitudinal vector bosons $W_{L}$
with other physical particles are the same as the S-matrix elements
of the theory where the $W_{L}$'s have been replaced by \emph{physical}
Goldstone bosons $(s^{\pm})$.
We are not going to get into details here; apart from the original literature
the reader is also referred to the  
articles~\cite{Bagger:1989fc,Shifman:2011ri,Marciano:2011gm,Jegerlehner:2011jm}.
%
%%%%%%%%%%%%%%%%
\begin{figure}[t] %  figure placement: here, top, bottom, or page
   \centering
   \includegraphics[width=4in]{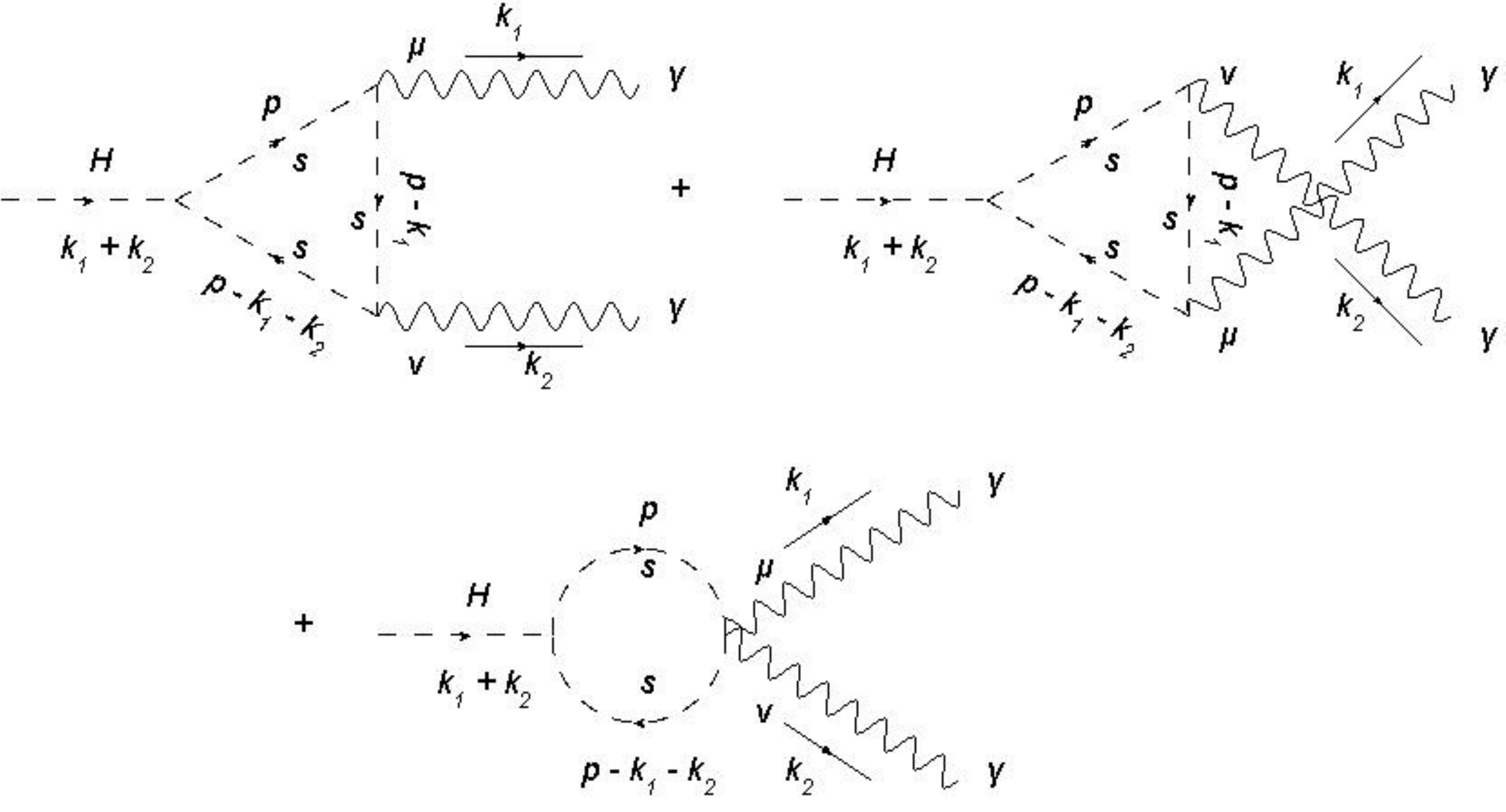} 
   \caption{Charged Goldstone boson contributions to $H\to \gamma\gamma$ in the limit of 
   $g\to 0$.}
    \label{fig2}
\end{figure}
%%%%%%%%%%%%%%%%%%%%%
%
Following \Ref{Bagger:1989fc}, within perturbation theory and in the limit of 
high energies, $m_{W}^{2}/s \to 0$,
GBET can be expressed with physics in two different limits of the theory: (a)
$g^{2}/\lambda_{H} \to 0$, or (b) $m_{H}^{2}/s \to 0$.

The limit (b) is irrelevant\footnote{The limit (b) simply says that matrix elements
for the theory which contains the 
physical $W_{L}$'s and zero v.e.v  is equal to those 
produced by scattering of massless physical Goldstone bosons (instead of $W_{L}$'s)
 at high energies.
We have checked that  \eq{GBET} is satisfied in this limit.} 
for defining $\lambda$ in \eq{eq:2.12} so we completely focus
on the limit (a). It is very easy to see that, in the unitary gauge,
the $W_{L}$'s  do not  decouple\footnote{This is another
advantage of calculating in the unitary gauge. Note that here, 
the word ``decoupling'' is stated in a different context than is usually quoted following the 
Appelquist-Carazzone~\cite{Appelquist:1974tg} 
theorem, where particle masses circulating in loops are much heavier than the external ones.}

for vanishing gauge coupling $g$. 
Consider for example
the diagrams in Fig.~\ref{fig1}: there is always a $m_{W}^{2}$ from the $HWW$-vertex
that cancels another $m_{W}^{2}$ sitting in the denominator of  the longitudinal part  
for the internal W-boson propagator expression written in the unitary gauge. 
So, as it was already noted in the paragraph below \eq{ma2},
in the limit $g\to 0$
there are remaining non-decoupled terms. Unfortunately,  these effects
may be obscured or misjudged by the regularisation method needed to handle divergent, 
intermediate, loop integrals.
This is exactly what happens  here when trying to calculate $\lambda$ directly from its
ambiguous form (\ref{lam}). 
On the other hand however, at the exact 
$g= 0$, with fixed v.e.v $v$ and Higgs quartic coupling $\lambda_{H}$,
\eq{GBET} suggests that the Goldstone bosons ($s^{\pm}$) should reappear at the 
physical spectrum of the theory while the longitudinal components of $W$'s become unphysical. 
At this limit, the SM is a spontaneously 
broken global $SU(2)_{L}\times U(1)_{Y}$-symmetry that couples, minimally, 
to electromagnetism. 
The interactions between the Higgs  and photon with the Goldstone bosons are simply 
those of a spontaneously broken scalar QED with $U(1)_{\rm em}$,
%%%%%%%%%%%%%%%
\begin{equation}
H\, s^{+}\, s^{-}  \;:\; -\frac{i  m_{H}^{2}}{v}\quad , \quad
\gamma \, s^{+}(p_{1}) \, s^{-}(p_{2}) \;:\; -i e (p_{1} + p_{2})^{\mu} \quad , \quad
\gamma \, \gamma \, s^{+} \, s^{-} \;:\;  2 i e^{2} g^{\mu\nu} \;. \label{frs}
\end{equation}
%%%%%%%%%%%%%%%%%%
%

Armed with these Feynman rules 
we calculate the diagrams in Fig.~\ref{fig2}. By doing so, we introduce again
three momentum variable shift vectors, one for each diagram, exactly in the same way
we did for the calculation of the diagrams in Fig.~\ref{fig1}. 
The Lorentz structure of the amplitude is completely analogous to \eq{gam2} with 
$\mathcal{M}_{1,2}\to \mathcal{M}_{1,2 (\rm GBET)}$, 
but now due to the scalar propagators,
% contributions to $\mathcal{M}_{2 (\rm GBET)}$, 
 the superficial degree of divergence, 
 for diagrams contributing to $\mathcal{M}_{2 (\rm GBET)}$, 
 is $D=-2$. Hence,
 all integrals involved in $\mathcal{M}_{2 (\rm GBET)}$ are finite
 and in addition, they are independent of any momentum integration
 shift vector variable.   As a consequence, $\mathcal{M}_{2 (\rm GBET)}$ is well defined, calculable,
 independent of any regularisation method,
 and  at the limit of $g\to 0$ (or 
$\beta = 4 m_{W}^{2}/m_{H}^{2}\to 0$)  is 
%%%%%%%%%%%%%%%%
\begin{equation}
\mathcal{M}_{2 (\rm GBET)} =
 - \frac{2 e^{2} g}{ (4\pi)^{2} m_{W} }
\;, \qquad \beta \to 0 \;. \label{eq:2.15}
\end{equation}
%%%%%%%%%%%%%%%%%
By equating \eq{eq:2.12} (in the limit $\beta \to 0$) and \eq{eq:2.15}
which represent the l.h.s and r.h.s 
of the GBET condition (\ref{GBET}), respectively, we find
%%%%%%%%%%%%%
\begin{equation}
\lambda = -1 \;. \label{216}
\end{equation}
%%%%%%%%%%%%%%%
%This is the value for $\lambda$ consistent with the two aspects,  (a) and (b), of GBET. 
%
This value agrees with dimensional regularization~\cite{'tHooft:1972fi,Collins} 
in the limit $d\to 4$  [see \eq{lamdef}].
The final form of the $\mathcal{M}_{2}$  in \eq{gam2} is
%%%%%%%%%%%%%%
\begin{equation}
\mathcal{M}_{2} \ = \ - \frac{e^{2} g }{(4 \pi)^{2} \: m_{W}} \: \biggl 
\{2  + \biggl [ 3\: \beta + 3 \: \beta\: (2-\beta)\: f(\beta) \biggr ] \biggr \} \;,
\label{a2}
\end{equation}
%%%%%%%%%%%%%%
with $\beta$, and $f(\beta)$ defined in \eq{bfb}. 

To complete the picture there is still the coefficient $\mathcal{M}_{1}$ in \eq{gam2}
to be calculated. 
Naive power counting says that this is by two powers more divergent than $\mathcal{M}_{2}$
and, in general, undetermined. It can be fixed however by 
using quantum gauge invariance i.e., conservation of charge, for the $U(1)_{\rm em}$,
%%%%%%%%%%%%%%%%%%%
\begin{equation}
k_{1\mu} \mathcal{M}^{\mu\nu} = 0\;, \quad k_{2\nu} \mathcal{M}^{\mu\nu} = 0 \;, \quad
k_{1}^{2} = k_{2}^{2} =0  \;,
\end{equation}
%%%%%%%%%%%%%%%%%%
and thus from \eq{gam2}, 
%%%%%%%%%%%%%%%%%%%
\begin{equation}
\mathcal{M}_{1} = - (k_{1}\cdot k_{2})\, \mathcal{M}_{2}   
%= -\mathcal{M}_{2} \,\left( \frac{m_{H}^{2}}{2} \right)
\;. \label{a12}
\end{equation} 
%%%%%%%%%%%%%%%%%
\Eq{a12} is substituted to \eq{gam2} with $\mathcal{M}_{2}$ read by \eq{a2}.
This is exactly the same result for the $W$-boson contribution to 
$H\to \gamma \gamma$ amplitude,
that has been obtained  in~\Refs{Ellis:1975ap,Ioffe:1976sd,
Shifman:1979eb,Gavela:1981ri,Marciano:2011gm} 
using dimensional regularisation in $R_{\xi}$-gauges. 
%%%%%%%%%%%%%%%%%%%%%%%%%%%%%%%%%%%

It is interesting here to note the result from the explicit algebraic manipulation of $\mathcal{M}_{1}$
in the unitary gauge and check the validity of gauge invariance [\eq{a12}].  
Exactly as for $\mathcal{M}_{2}$,
the condition $a=b=c$ for the arbitrary vectors given in \eq{con}
is crucial in reducing the divergence 
of $\mathcal{M}_{1}$  down to a logarithmic one [see expression \eq{eq:A11}].
In $d$-dimensions  the expression for $\mathcal{M}_{1}$ is finally independent of
any arbitrary vector and, up to 
a proportionality factor, reads:
%\begin{eqnarray}
%\mathcal{M}_{1} \sim \frac{4}{m_{W}^{2}}\int_{0}^{1}dx\int_{0}^{1-x}dy\int\frac{d^{4}\ell}{(2
%\pi)^{4}}& \bigg \{&\frac{\bigl[4(\ell\cdot k_{1})(\ell\cdot k_{2})
%-\ell^{2}(k_{1}\cdot
%k_{2} )\bigr]}{(l^{2}-\Delta)^{3}}+  \nonumber \\
%&+& \frac{\bigl[3\,m_{W}^{4}-3 \,m_{W}^{2}\, m_{H}^{2}-3\,
%x(x+y-1)m_{W}^{2}\,m_{H}^{2} \bigr]}{(l^{2}-\Delta)^{3}}\biggr \}
%\end{eqnarray}
\begin{align}
\mathcal{M}_{1} \sim  4\: \int_{0}^{1}dx\int_{0}^{1-x}dy\int\frac{d^{d}\ell}{(2
\pi)^{d}} &\biggl  \{\frac{4(\ell\cdot k_{1})(\ell\cdot k_{2})
+2(\frac{2}{d}-1)\ell^{2}(k_{1}\cdot
k_{2} )+(\frac{d-1}{d})(4-d)\,\ell^{2}\,m_{W}^{2}}{(\ell^{2}-\Delta)^{3}}+  \nonumber \\[3mm]
&+ \frac{(d-1)\,m_{W}^{4}-3 \,m_{W}^{2}\, m_{H}^{2}+(1-d)\,
x(x+y-1)m_{W}^{2}\,m_{H}^{2} }{(\ell^{2}-\Delta)^{3}}\bigg\}\;. \label{m1}
\end{align}
%%%%%%%%%%%%%%%%%%%%%%%%%%%%%%%%%
Clearly the first integral in \eq{m1} is ill-defined in four dimensions.
If however, we  insist in doing the calculation of \eq{m1} 
in $d=4$ with symmetric integration, like in 
\Refs{Gastmans:2011ks,Gastmans:2011wh}, we find that gauge invariance [\eq{a12}] is
\emph{not} satisfied. This  is of course unacceptable. 
By going a little bit deeper, gauge invariance is lost because of  
the term proportional to $4-d$ in \eq{m1}  which vanishes when $d=4$.
Quite the contrary in DR, this term 
results in a non-zero contribution when $m_{W}\ne 0$, since the (log divergent) integral 
in front of $(4-d)$ contains a simple pole at $d=4$. 
This changes the final result and renders \eqss{a2}{m1}{a12}
consistent, only if $\lambda = -1$. This outcome  is in agreement with \Ref{Marciano:2011gm}.

Few remarks are worth mentioning here. 
Had we started first calculating $\mathcal{M}_{1}$, there would be no possibility
of defining unambiguously $\lambda$ without using a gauge invariant regulator: 
the $g_{\mu\nu}$ part
of the amplitude at $g\to 0$ involving Goldstone bosons [see diagrams fig.~\ref{fig2}]
is not well defined - an integral as the one in \eq{mann2} appears again.
Another remark is that the same expressions for the coefficients  $\mathcal{A}_{ij}$
displayed in \ref{sec:appA}
 in the unitary gauge, appear also when one exploits the  $R_{\xi}$-gauge.
In the latter there are in addition $\xi$-dependent terms~\cite{Marciano:2011gm}
 that vanish in the end from unphysical scalar 
contributions.  Therefore, the logarithmic  ambiguity in \eq{mann2}, 
found here in the unitary gauge, is similar in every other  gauge.
%However, it is possible,
 %for a specific gauge fixing choice i.e., $\xi=1$~\cite{Piccinini:2011az}, to obtain
%a well defined $\mathcal{M}_{2}$ coefficient 
%but this should only be accidental. 
%
%Second, if I apply the result of 2.12 of our draft to the first term of
%expression (1) I find that this is equal to zero in 4-dimensions.
%If I repeat the same procedure,but taking \omega->2 (from the same expression
%2.12) I find that the result is 1/2 of the result that I find when use DR in
%expression (2) and this is another problem.

%%%%%%%%% NEW TEXT v4 %%%%%%%%%%%%%%%%
%
The ambiguity of the integral in \eq{mann2} has been discussed 
by many articles in the recent literature. \Rrefs{Huang:2011yf,Shao:2011wx}
have used gauge-invariant regulators al\'a Pauli-Villars. Most notably, 
Piccinini {\it et.al}~\cite{Piccinini:2011az} showed  that the unitary gauge with a cut-off regularisation 
scheme turns out to be non-predictive: new physics input, along the lines of 
\Ref{Jackiw:1999qq} also followed  in our
article, is needed.  As already noted, Marciano {\it et.al}~\cite{Marciano:2011gm} were the first 
to make the calculation of $h\to \gamma\gamma$ in unitary gauge with DR. 
Furthermore, 
\Refs{Shifman:2011ri,Jegerlehner:2011jm,Huang:2011yf,
Piccinini:2011az,Marciano:2011gm} showed that the ``decoupling limit'' $m_{W}/m_{H}\to 0$
{\emph must} hold  because of the GBET.   All these articles together with the
one at hand conclude that the result in \Refs{Gastmans:2011ks,Gastmans:2011wh}
is incorrect.
Furthermore, the aim of our paper is not to redo the calculation 
in the unitary gauge with DR as in \Ref{Marciano:2011gm}; this is only a
by-product of our analysis. On the contrary,
we want to clarify subtle issues related to this calculation
in the unitary gauge and in $d=4$ raised in part by \Refs{Gastmans:2011ks,Gastmans:2011wh}. 
We find by working strictly in $d=4$, using arbitrary vectors, 
that divergencies (up to the 6th power) are reduced  down to 
logarithmic ones. This is a new result that is not obvious
when working in the unitary gauge and cannot be seen 
when using dimensional regularisation.

%%%%%%%%%%%%%%%%%%%%%%%
\section{Four Dimensional Regularization (FDR)}
\label{sec:fdr}
%%%%%%%%%%%%%%%%%%%

So far we have proposed a regularization scheme which is four-dimensional 
and uses the basic symmetries and underlying physics of the SM.
However, in more complicated models or observables with more parameters
to adjust,  such a scheme can become cumbersome.  For example, it is not always
obvious which physics argument will  fix undefined integrals.    

Very recently, R.~Pittau~\cite{Pittau:2012zd} proposed a scheme that is fairly 
easy to handle  and, to the best of our knoweledge, is the 
closest to four dimensional calculations, thereby coined 
four-dimensional regularisation/renormalization scheme
or just FDR. According to this scheme, infinite bubble graph contributions,
i.e., large loop momenta contributions that do not depend upon external momenta,
are absorbed into the shift of the vacuum while the remaining finite corrections 
are calculable in four-dimensions in addition to being Lorentz and gauge invariant.

We have applied FDR into the calculation of the $H\to \gamma\gamma$
amplitude and found agreement with our physics approach and with DR results. 
In FDR one introduces an arbitrary scale $\mu$ which is considered to be
much smaller than internal momenta and particle masses in loops. Self contracted loop 
momenta quantities like $\ell^{2}$ become $\bar{\ell}^{2}=\ell^{2} - \mu^{2}$, while 
for gauge invariance to hold, vector momenta, $p^{\mu}$, remain untouched.
For example the integral of \eq{mann2} becomes,
%%%%%%%%%%%%%%%
 \begin{equation}
 \int [d^{4}\ell] \; \frac{\bar{\ell}^{2}\, g_{\mu\nu} - 
 4\,\ell_{\mu}\, \ell_{\nu}}{\bar{D}^{3}} \ = 
 \int [d^{4}\ell] \; \frac{-\mu^{2}}{\bar{D}^{3}} \: g_{\mu\nu}\;,
 \label{mann3}
 \end{equation}
 %%%%%%%%%%%%%%%%%%
where $\bar{D} = (\bar{\ell}^{2} - \Delta)$ and $[d^{4}\ell]$ stands for integration
over $d^{4}\ell$, dropping all divergent terms from the integrand (see below) and 
 taking the limit $\mu \to 0$. In going from l.h.s to r.h.s of
\eq{mann3} the symmetry property $\ell_{\mu}\ell_{\nu} = g_{\mu\nu}\, \ell^{2}/4$ 
has been used in four
dimensions. Then, using the partial  fractions identity,
%%%%%%%%%%%%%%%
\begin{eqnarray}
\frac{1}{\bar{D}^{3}} = \biggl [\frac{1}{\bar{\ell}^{6}} \biggr ]  + 
\Delta \: \left (\frac{1}{\bar{D}^{3} \bar{\ell}^{2}}  + \frac{1}{\bar{D}^{2} \bar{\ell}^{4}} 
+ \frac{1}{\bar{D} \bar{\ell}^{6}} \right )\;,
\end{eqnarray}
%%%%%%%%%%%%%%% 
the term in square bracket is recognised as divergent and therefore removed, and 
integrating  the r.h.s of \eq{mann3} over $[d^{4}\ell]$  one obtains
%%%%%%%%%%%%%%%%
\begin{eqnarray}
  \int [d^{4}\ell] \; \frac{-\mu^{2}}{\bar{D}^{3}} \equiv -\Delta \lim_{\mu\to 0} \mu^{2}
  \int d^{4}\ell \left (\frac{1}{\bar{D}^{3} \bar{\ell}^{2}}  + \frac{1}{\bar{D}^{2} \bar{\ell}^{4}} 
+ \frac{1}{\bar{D} \bar{\ell}^{6}} \right ) = -\frac{i\pi^{2}}{2} \;,
\end{eqnarray}
%%%%%%%%%%%%%%%
i.e., exactly the same result as in DR which eventually leads to $\lambda=-1$ 
consistent with gauge invariance
and GBET. What in fact FDR scheme does, is to restate 
the correct DR answer through the regulator $\mu^{2}$ 
keeping \eq{mann2} correct in $d=4$. We therefore understand that the constant ($\beta$-independent) term of \eq{a2} in FDR arises from the fact that the arbitrary scale, 
$\mu^{2}$, \emph{must}
 disappear from physical observables.   

%%%%%%%%%%%%%%%%%%%%%%%
\section{Discussion}
\label{discussion}
%%%%%%%%%%%%%%%%%%%

It is evident that our calculation  for the amplitude incorporates two physical
inputs: one is the conservation of charge and the other is the equivalence 
theorem. They are both direct consequences of the gauge invariance of the 
underlying physical theory.
The first one is experimentally indisputable 
while the second one is theoretical\footnote{This is not entirely correct. There 
is of course  the high energy behaviour of $e^{+}e^{-}\to W^{+} W^{-}$ found
at LEP~\cite{Abdallah:2010zj} consistent with the GBET.} 
and has been proven in \Ref{Tiktopoulos} that is valid 
in any spontaneously broken renormalizable theory, like for example the SM. 
One may think however that there is a loophole in our use of this  second argument:
so far, and, to our knowledge,  the replacement of the $W$-bosons with Goldstone bosons
at high energies  has been proven  to be valid only for external 
$W$-bosons~\cite{Chanowitz:1985hj,Gounaris:1986cr,Bagger:1989fc} 
and \emph{not} for internal ones which is the case exploited here. 
%for diagrams in Fig.~\ref{fig2}. 
Although it has been tested in
several phenomenological examples~\cite{Dawson:1989up}, a formal, to all orders, proof  is still
missing. Although this may be true, it is difficult to argue against  the validity of decoupling limit
$g\to 0$ (with fixed v.e.v and Higgs quartic coupling) 
discussed in the paragraph above \eq{frs}. 
  
%This brings us to the next question : 
%can we define $\lambda$ by any  other ``physics''  or ``self consistency''
%argument?   YES low-energy Higgs theorem or trace anomaly. {\bf say more}

Is there another physics context from which one can define $\lambda$?
One possibility is to exploit the 
low  energy 
Higgs theorem~\cite{Ellis:1975ap,Vainshtein:1980ea,Kniehl:1995tn,Pilaftsis:1997fe} instead.
Although this may serve as a consistency check, and indeed is compatible with $\lambda=  -1$,
we cannot use it to define $\lambda$. The reason here is threefold: first, when treating the
Higgs field as an external background  field with zero momentum one needs to take 
partial derivative w.r.t $m_{W}$ of the 2-point photon vacuum polarization amplitude, 
$\Pi_{\gamma\gamma}(q^{2})$. 
The later, is notoriously difficult, if meaningful, to be calculated in the unitary gauge.
Second, according to \Ref{Ellis:1975ap}, we know that to the lowest order in weak coupling,
the amplitude for the process $\braket{\gamma\gamma}{H}$ is proportional to
$\bra{\gamma\gamma}\Theta_{\mu}^{\, \mu}\ket{0}$ where 
$\Theta_{\mu}^{\, \mu} = 2 m_{W}^{2} W^{+} W^{-} + ...$
is the improved energy momentum tensor~\cite{Callan:1970ze}. However, the calculation of 
$\bra{\gamma\gamma}\Theta_{\mu}^{\, \mu}\ket{0}$ goes through the 
same steps as for the calculation for the $H\to \gamma\gamma$ amplitude
and therefore involves the same ambiguity for calculating $\lambda$. 
Third, one could examine the $W$-contribution to $H\to \gamma\gamma$ within
the dispersion relation approach. It can be shown~\cite{Horejsi:1996ak} that the non-vanishing
limit at $g_{W}\to 0$ is due to a finite subtraction induced by the corresponding trace
anomaly~\cite{Adler:1976zt}. However, in order to calculate unambigiously this finite piece, 
one has to make full use of  a (physical) boundary condition of the theory.
%such as, for example, the GBET. 
 %In conclusion, we find that an unambiguous 4-dimensional 
 %calculation in unitary gauge for the $H\to \gamma\gamma$ amplitude
%must be performed within a physical setup. We have demonstrated this statement 
%by exploiting GBET.  

As a final remark, suppose that we did not know DR and wanted to calculate a
 certain observable in 4-dimensions. In this observable we encounter singularities
 i.e.,  undefined and undetermined integrals. Then we use physics
 arguments to fix these ambiguities. However, we can always question
 whether we are using the right physics set up or not. In that sense
 the final judgement should come from the experiment.
Therefore it may be not academic to ask whether 
LHC could see the difference between $\lambda=-1$ and $\lambda =0$?
Setting the SM Higgs mass $m_{H} = 125$ GeV, and
including  the top-loop contribution, we find 
%%%%%%%%%%%%%
\begin{equation}
\frac{\mathrm{Br}(H\to \gamma\gamma,\, \lambda=0)}{\mathrm{Br}(H\to 
\gamma\gamma,\, \lambda=-1)} \ \approx \ 0.46 \;.
\end{equation}
%%%%%%%%%%%%%%%
This is certainly within LHC's sensitivity for 14 TeV c.m energy and luminosity of $30\, fb^{-1}$.
(see for example Fig.~3 in \Ref{Klute:2012pu}). In fact,  the recent observation 
by LHC experiments~\cite{ATLAS:2012gk,CMS:2012gu} indicates
a value 
$\frac{\mathrm{Br}(H\to \gamma\gamma,\, \mathrm{(exp)})}{\mathrm{Br}(H\to 
\gamma\gamma, \, \lambda=-1)} \ = \ 1.6 \pm 0.3$~\cite{Giardino:2012dp}
 which highly disfavours the case 
 $\lambda=0$ by almost four standard deviations. 
 We can turn this around and state that this is  an indirect hint
towards the validity of the equivalence theorem. 

%In fact, the best fit 
%analysis~\cite{Carmi:2012yp} to current 
%(2011-2012a) Tevatron  and LHC  data  seems to disfavour the 
%case $\lambda =0$.}.

%%%%%%%%%%%%%%%%%%%%%%%%%%%%
\section{Conclusions}
\label{conclusions}

In this work we review the $W$-gauge boson loop contribution to 
the $H\to \gamma\gamma$ amplitude in the unitary gauge. Our objective is to fix 
intermediate step indeterminacies arising from divergent diagrams by making full use of
physics at $d=4$ much in the same way as in the calculation of the
chiral anomaly triangle.

We anticipate a finite result for the loop induced 
$H\to \gamma \gamma$-amplitude  in the renormalizable SM.
Therefore the amplitude has to be independent of any shifting momentum variables
we have originally introduced. But finite or even log divergent integrals are independent
of these vectors, so  the vectors  have to be accompanied only by infinite contributions, 
if at all. Therefore,
infinities and arbitrary vectors are eliminated altogether by a certain combination among
them [see \eq{con}].

The whole calculation in the unitary gauge boils down
to  a logarithmically divergent integral (\ref{lam}). We find that, 
this  integral results in two different values depending on whether $d\to 4$ or $d=4$.
This is due to a surface term remaining at the exact $d=4$ case after the part-by-part 
integration in $d$-dimensions [see~\ref{sec:appB}]. 
To proceed, we identify this integral with an undefined parameter $\lambda$ [see \eq{mann2}].
This  parameter is then fixed unambiguously 
by assuming the validity of the Goldstone Boson Equivalence 
Theorem (GBET). 
Its value is consistent with DR in the limit $d\to 4$. 

In our calculation we are very careful not to perform shifting of integration variables
for highly divergent integrals by introducing three arbitrary momentum variable 
shift vectors straight from the beginning. 
Divergencies and arbitrariness 
from these unknown vectors are altogether removed, 
%by adopting a regularisation scheme 
leaving behind  a log-like divergent integral in $\mathcal{M}_{2}$ of \eq{gam2}.
This is defined 
by a  physical input taken from the GBET
and is connected to $\mathcal{M}_{1}$ by electromagnetic charge conservation.

As noted many times in the text, the key point  towards deriving an unambiguous amplitude
for $H\to \gamma\gamma$   in the unitary gauge  
is the limit of vanishing gauge couplings; this is an aspect of GBET [\eq{GBET}]. 
In this limit, the Goldstone boson
loop contributions to the coefficient $\mathcal{M}_{2}$ is finite, i.e.,  
independent of any regularisation
scheme.  

We also saw that DR (FDR), a regularisation
scheme introduced to maintain Ward Identities at intermediate steps of a calculation,
supports the 
GBET in the limit $d\to 4$ ($d=4$). 
On the contrary,  we find that,  performing the integrals in $d=4$
with symmetric integration is not a good choice because it leads 
to the violation of gauge invariance [see \eq{a12} and the discussion below].
The main reason is due to surface terms that are developed in exactly $d=4$ dimensions 
 [see discussion below \eq{lam} and \ref{sec:appB}]. 
The latter are axiomatically discarded in DR~\cite{'tHooft:1972fi,Collins}.
Another reason is the appearance of the $(d-4)$-term in the numerator of \eq{m1}.
%

%Another example of an observable calculated in unitary gauge and
%found consistent with 
%Although we are aware of other observables performed in unitary gauge, like
%for example  $Z\to \bar{t} c$ in \Ref{Axelrod:1982yc},
% we are  not aware of a systematic proof for   
% unambiguous loop results in this, non-renormalizable, gauge.  

In conclusion, 
the four-dimensional calculation of $H\to \gamma \gamma$ amplitude in the unitary gauge
is ambiguous without introduction of a physics input beyond gauge invariance. 
As we have demonstrated, this physics, which uniquely defines the amplitude, 
may arise from the Goldstone Boson Equivalence Theorem (GBET). 
This effectively proves that GBET comprises an additional 
important  pillar of the Standard Model dynamics.

\bigskip
%%%%%%%%%%%
{\bf Acknowledgments:}
We are grateful to Adrian Signer for illuminating discussions and critical
comments on the manuscript.
We would also like to thank   R. Pittau, R. Jackiw, F. Piccinini, 
P.~Kanti, S.~Martin and K.~Tamvakis for  discussions
and comments. We also 
thank C.~Coriano for bringing to our attention \Ref{Horejsi:1996ak}.
%
%This research has been co-financed by the European Union (European  
%Social Fund 8211; ESF) and Greek national funds through the  
%Operational Program "Education and Lifelong Learning" of the National  
%Strategic Reference Framework (NSRF) - Research Funding Program ARISTEIA.
%
This research Project  is co-financed by the European Union -  
European Social Fund (ESF) and National Sources, in the framework of the  
program ``ARISTEIA" of the ``Operational Program Education and Lifelong  
Learning" of the National Strategic Reference Framework (NSRF)  
2007-2013.
K.S. acknowledges full financial support from Greek State
Scholarships Foundation (I.K.Y).
%%%%%%%%%%%%%%%%%%%%%%%%%%%%%%%%
%\end{acknowledgments}

%\appendix

%\section{Formulae}

%\newpage
%  \section*{APPENDIX}  % use *-form to suppress numbering

%\begin{appendix}
%\appendix
%\section{Appendix}
%\renewcommand{\theequation}{A-\arabic{equation}}    
%\setcounter{equation}{0}  % reset counter     
% redefine the command that creates the section and equation no.
\renewcommand{\thesection}{Appendix~\Alph{section}}
\renewcommand{\theequation}{\Alph{section}.\arabic{equation}}

\setcounter{equation}{0}  % reset counter
\setcounter{section}{0}
\bigskip

%%%%%%%%%%%%%%%%%%%%%%%
\section{}
\label{sec:appA}

We append here the integrand expressions for the 
coefficients $\mathcal{A}_{ij}$ in \eq{mat}. 
The corresponding formula
for $\mathcal{A}_{11}$ is quite long and  is not included
here. It can be provided by the authors upon request.
Note that  the number of dimensions $d$  has been kept arbitrary throughout and
on-shell conditions for the external particles have been imposed.
%%%%%%%%%%%%%%%%%%%%%%%
\begin{eqnarray}
\mathcal{A}_{21}=\frac{1}{[(p+a)^{2}-m^{2}_{W}][(p+a-k_{1})^{2}-m^{2}_{W}][(p+a-k_{1}-k_{2})^{2}-m^{2}_{W}]}\times \nonumber \\
\bigg \{(4 \ d-6)\: m_{W}^{2}+ \bigg[3 \ (p+a)\cdot
(p+a)-5 \ (p+a)\cdot k_{1}-(p+a)\cdot k_{2}+2 \ m^{2}_{H} \bigg]+
\nonumber \\
+\frac{1}{m^{2}_{W}}\bigg[-((p+a)\cdot (p+a))^{2}+3 \ ((p+a)\cdot
k_{1})((p+a)\cdot(p+a))-2 \ ((p+a)\cdot k_{1})^{2}- \nonumber \\
-2
\ ((p+a)\cdot k_{1})((p+a)\cdot k_{2})+((p+a)\cdot
(p+a))((p+a)\cdot k_{2}) \bigg]\bigg\}\;, \nonumber \\
\end{eqnarray}
%%%%%%%%%%%%%%%%%%%%%%%%%
%
%%%%%%%%%%%%%%%%%%%%%%
\begin{eqnarray}
\mathcal{A}_{22}=\frac{1}{[(p+b)^{2}-m^{2}_{W}][(p+b-k_{2})^{2}-m^{2}_{W}][(p+b-k_{1}-k_{2})^{2}-m^{2}_{W}]}\times \nonumber \\
\bigg \{(4 \ d-6)\: m_{W}^{2} +  \bigg[3 \ (p+b)\cdot
(p+b)-5 \ (p+b)\cdot k_{2}-(p+b)\cdot k_{1}+2 \ m^{2}_{H} \bigg]+
\nonumber \\
+\frac{1}{m^{2}_{W}}\bigg[-((p+b)\cdot (p+b))^{2}+3 \ ((p+b)\cdot
k_{2})((p+b)\cdot(p+b))-2 \ ((p+b)\cdot k_{2})^{2}- \nonumber \\
-2 \ ((p+b)\cdot k_{1})((p+a)\cdot k_{2})+((p+b)\cdot
(p+b))((p+b)\cdot k_{1}) \bigg]\bigg\}\;, \nonumber \\
\end{eqnarray}
%%%%%%%%%%%%%%%%%%%%%%
\begin{eqnarray}
\mathcal{A}_{23}=\frac{-1}{[(p+c)^{2}-m^{2}_{W}][(p+c-k_{1}-k_{2})^{2}-m^{2}_{W}]}
 \times \nonumber \\
  \bigg \{4+\frac{2}{m^{2}_{W}} \bigg[-((p+c)\cdot(p+c))+(p+c)\cdot k_{1}+(p+c)\cdot k_{2} \bigg]\bigg
 \}\;, \nonumber \\
\end{eqnarray}
%%%%%%%%%%%%%%%%%%%%%%%%%%%%
\begin{eqnarray}
\mathcal{A}_{31}=\frac{1}{[(p+a)^{2}-m^{2}_{W}][(p+a-k_{1})^{2}-m^{2}_{W}][(p+a-k_{1}-k_{2})^{2}-m^{2}_{W}]}\times \nonumber \\
\bigg \{(7-4 \ d)\: m_{W}^{2} - \bigg[4 \ (p+a)\cdot (p+a)-7
\ (p+a)\cdot k_{1}+3 \ (p+a)\cdot k_{2} \bigg]+
\nonumber \\
+\frac{1}{m^{2}_{W}}\bigg[((p+a)\cdot (p+a))^{2}-3 \ ((p+a)\cdot
k_{1})((p+a)\cdot(p+a))+2 \ ((p+a)\cdot k_{1})^{2}+ \nonumber \\
+2 \ ((p+a)\cdot k_{1})((p+a)\cdot k_{2})-((p+a)\cdot
(p+a))((p+a)\cdot k_{2}) \bigg]\bigg\}\;, \nonumber \\
\end{eqnarray}
%%%%%%%%%%%%%%%%%%%%%%%%%%%%%%%%%%%%%%%%%%%%%%%%%
%%%%%%%%%%%%%%%%%%%%%%%%%%%%%%
\begin{eqnarray}
\mathcal{A}_{32}=\frac{-1}{[(p+b)^{2}-m^{2}_{W}][(p+b-k_{2})^{2}-m^{2}_{W}][(p+b-k_{1}-k_{2})^{2}-m^{2}_{W}]}\times \nonumber \\
\bigg \{m_{W}^{2} + \bigg[-(p+b)\cdot (p+b)+6 \
(p+b)\cdot k_{2}\bigg ]\bigg \} \;, \nonumber \\ 
\end{eqnarray}
%%%%%%%%%%%%%%%%%%%%%%%%%%%%%%%%%%%%%%%%%%%%%%%%%%
\begin{eqnarray}
\mathcal{A}_{33}=\frac{1}{[(p+c)^{2}-m^{2}_{W}][(p+c-k_{1}-k_{2})^{2}-m^{2}_{W}]}
 \times \nonumber \\ \bigg \{2-\frac{1}{m^{2}_{W}} \bigg[(p+c)\cdot(p+c)-(p+c)\cdot k_{1}-(p+c)\cdot k_{2} \bigg]\bigg
 \}\;, \nonumber \\
 \end{eqnarray}
 %%%%%%%%%%%%%%%%%%%%%%%%%%%%%%%%%%%%%%%%%%%%%%%%%
 \begin{eqnarray}
 \mathcal{A}_{41}=\frac{-1}{[(p+a)^{2}-m^{2}_{W}][(p+a-k_{1})^{2}-m^{2}_{W}][(p+a-k_{1}-k_{2})^{2}-m^{2}_{W}]}\times \nonumber \\
\bigg \{m_{W}^{2}+ \bigg[-(p+a)\cdot (p+a)+6 \
(p+a)\cdot k_{1}\bigg ]\bigg \}\;, \nonumber \\
\end{eqnarray}
%%%%%%%%%%%%%%%%%%%%%%%%%%%%%%%%%%%%%%%%%%
\begin{eqnarray}
\mathcal{A}_{42}=\frac{1}{[(p+b)^{2}-m^{2}_{W}][(p+b-k_{2})^{2}-m^{2}_{W}][(p+b-k_{1}-k_{2})^{2}-m^{2}_{W}]}\times \nonumber \\
\bigg \{(7-4 \ d)\: m_{W}^{2} - \bigg[4 \ (p+b)\cdot (p+b)-7
\ (p+b)\cdot k_{2}+3 \ (p+b)\cdot k_{1} \bigg]+
\nonumber \\
+\frac{1}{m^{2}_{W}}\bigg[((p+b)\cdot (p+b))^{2}-3 \ ((p+b)\cdot
k_{2})((p+b)\cdot(p+b))+2 \ ((p+b)\cdot k_{2})^{2}+ \nonumber \\
+2 \ ((p+b)\cdot k_{1})((p+b)\cdot k_{2})-((p+b)\cdot
(p+b))((p+b)\cdot k_{1}) \bigg]\bigg\}\;, \nonumber \\
\end{eqnarray}
%%%%%%%%%%%%%%%%%%%%%%%%%%%%%%%%
\begin{eqnarray}
\mathcal{A}_{43} \ = \ \mathcal{A}_{33}\;,   
\end{eqnarray}
%%%%%%%%%%%%%%%%%%%%%%%%%%%%%%%%
%%%%%%%%%%%%%%%%%%%%%%%%%%%%%%%%%%%%%%%%%%%
\begin{eqnarray}
 \mathcal{A}_{51}=\frac{1}{[(p+a)^{2}-m^{2}_{W}][(p+a-k_{1})^{2}-m^{2}_{W}][(p+a-k_{1}-k_{2})^{2}-m^{2}_{W}]}\times \nonumber \\
\bigg \{5\: m_{W}^{2}+ \bigg[3 \ (p+a)\cdot (p+a)-2 \
(p+a)\cdot k_{1}\bigg ]\bigg \}+ \nonumber \\
+\frac{1}{[(p+b)^{2}-m^{2}_{W}][(p+b-k_{2})^{2}-m^{2}_{W}][(p+b-k_{1}-k_{2})^{2}-m^{2}_{W}]}\times \nonumber \\
\bigg \{5\: m_{W}^{2}+ \bigg[3 \ (p+b)\cdot (p+b)-2 \
(p+b)\cdot k_{2}\bigg ]\bigg \}- \nonumber \\
-\frac{2}{[(p+c)^{2}-m^{2}_{W}][(p+c-k_{1}-k_{2})^{2}-m^{2}_{W}]} \;.
 %\times \bigg(\frac{2}{m^{2}_{W}}\bigg)\;.
 %%%%%%%%%%%%%%%%%%%%%%%%%%%%%%%%%%%%%%%%%%%%%
\end{eqnarray}
%%%%%%%%%%%%%%%%%%%%%%%%%%%%%%%%%%%%%%%%%%
It is straightforward, but long and tedious, 
to show that after implementing the condition (\ref{con})  
to coefficients in 
eqs.(A.1)-(A.10) we arrive 
at \eq{M1} which is \emph{at the most logarithmically 
divergent.} 

For complementarity reasons, it is useful in deriving \eq{m1} 
to present the expression for the coefficient 
$\mathcal{A}_{11}$ after the imposition of the arbitrary vector relation \eq{con}:
%%%%%%%%%%%%%%%%%%%%%%%%%%%%%%%%
\begin{eqnarray}
\mathcal{A}_{11}=\frac{1}{[(p+a)^2-m_{W}^{2}][(p+a-k_{1})^{2}-m_{W}^{2}][(p+a-k_{1}-k_{2})^{2}-m_{W}^{2}]}\nonumber \\
\bigg \{\bigg ((p+a-k_{1})^{2}-m_{W}^{2}\bigg)\, (1-d)\, m_{W}^{2}+\nonumber \\
+\, 4 [(p+a)\cdot k_{1}][(p+a)\cdot k_{2}]-[3
m_{W}^{2}+(p+a)^{2}]\,m_{H}^{2} \bigg \}+\nonumber \\
+\frac{1}{[(p+a)^2-m_{W}^{2}][(p+a-k_{2})^{2}-m_{W}^{2}][(p+a-k_{1}-k_{2})^{2}-m_{W}^{2}]}\nonumber \\
\bigg \{ \bigg ((p+a-k_{2})^{2}-m_{W}^{2}\bigg)\,(1-d) \, m_{W}^{2}\ +\nonumber \\
+\, 4 [(p+a)\cdot k_{1}][(p+a)\cdot k_{2}]-[3
m_{W}^{2}+(p+a)^{2}]\, m_{H}^{2} \bigg \}\;. \label{eq:A11}
\end{eqnarray}
%%%%%%%%%%%%%%%%%%%%%%%%%%
This integrand 
expression, under $\int d^{4}p$, is obviously \emph{at the most logarithmically divergent}.

%%%%%%%%%%%%%%%%%%%%%%%%%%%%%%%%%%%
%\end{appendix}
%%%%%%%%%%%%%%%%%%%%%%%%

\medskip

%%%%%%%%%%%%%%%%%%%%%%%%%%
%\begin{appendix}
%\appendix
\setcounter{equation}{0}  % reset counter
\section{Dimensional Regularization and the surface term}
\label{sec:appB}
%\renewcommand{\theequation}{B-\arabic{equation}}    
%\setcounter{equation}{0}  % reset counter     
%%%%%%%%%%%%%%%%%%%%%%%
%%%%%%%%%%%%%%%%%%%%
%Within dimensional regularisation,  
We would like to examine the surface terms arising in $d=4$ when calculating 
 the integral on the l.h.s of \eq{mann2}. This integral  after Wick rotation
into Euclidean space, reads 
%%%%%%%%%%%%%%%
 \begin{equation}
   i\,  \int d^{2\omega}\ell \; \frac{\ell^{2}\, g_{\mu\nu} - 
 4\,\ell_{\mu}\, \ell_{\nu}}{(\ell^{2} + \Delta)^{3}}  \label{integ} \;,
 \end{equation}
 %%%%%%%%%%%%%%%%%%
where $\ell \equiv \ell_{E}$ and drop for clarity the subscript $E$  from now on. 
We follow very closely 
't Hooft and Veltman's seminal paper in \Ref{'tHooft:1972fi}.
In our calculation for a physical process 
we should notice first that $\ell_{\mu}, \ell_{\nu}$ are strictly 4-vectors since
they are contracted with physical external momenta $k_{1,2}^{\mu}$ or $k_{1,2}^{\nu}$.
On the other hand, the loop momentum $\ell$ in $\ell^{2}$ has components  in all, $d=2\omega$, dimensions.
We write $\ell$ as a sum of a vector $\ell_{\|}$ 
which has non-zero components in dimensions $0,1,2,3$ 
and a
vector $\ell_{\bot}$ which has nonzero components in $(2\omega-4)$-dimensions,
%%%%%%%%%%%%%%%
\begin{equation}
\ell \ = \ \ell_{\|} + \ell_{\bot} \;.
\end{equation}
%%%%%%%%%%%%%%%%
With this definition, the integral (\ref{integ}) reduces to 
%%%%%%%%%%%%%%%
 \begin{equation}
i \int d^{2\omega}\ell \; \frac{\ell_{\bot}^{2}\, g_{\mu\nu}}{(\ell^{2} + \Delta)^{3}}  \label{integ2} \;,
 \end{equation}
 %%%%%%%%%%%%%%%%%%
where the $\ell_{\|}$ components in the numerator of (\ref{integ}) vanish thanks to 
 symmetric integration formula, 
 $\ell_{\|}^{\mu} \ell_{\|}^{\nu} \to \frac{1}{4} \ell_{\|}^{2} g^{\mu\nu}$.  In order not to
 carry  the $g_{\mu\nu}$ in all formulae below we just concentrate on the integral 
%%%%%%%%%%%%%%%
 \begin{equation}
\mathcal{I} \
\equiv \ i \int d^{2\omega}\ell \; \frac{\ell_{\bot}^{2}}{(\ell^{2} + \Delta)^{3}}  = i
\int d^{4}\ell_{\|} \int d^{2\omega - 4} \ell_{\bot} \; \frac{\ell_{\bot}^{2}}{(\ell_{\|}^{2} + 
\ell_{\bot}^{2}+ \Delta)^{3}}  \label{integ3} \;.
 \end{equation}
%%%%%%%%%%%%%%%%%%%%%%%%%%%%%%
Integrating over the extra dimensional solid angle $d\Omega_{2\omega -4}$ we arrive
at 
%%%%%%%%%%%%%%%
\begin{equation}
\mathcal{I} = \frac{2\, i\, \pi^{\omega -2}}{\Gamma(\omega - 2)} \int d^{4}\ell_{\|} \int_{0}^{\infty} 
d L \, \frac{L^{2\omega -3}}{(\ell_{\|}^{2} + L^{2} + \Delta)^{3}} \;,
\label{b5}
\end{equation}
%%%%%%%%%%%%%% 
where $\Gamma(x)$ is the Euler $\Gamma$-function and $L$ is the length of the
 $\ell_{\bot}$ vector.
This integral is UV divergent for 
$\omega \ge 2$ and IR divergent for $\omega \le 1$. Therefore, the region of convergence,
$1 < \omega < 2$, 
is finite  but it does not yet include the point $\omega = 2$. In order to 
enlarge the region of convergence to include $\omega=2$
one has to analytically continue $\mathcal{I}$ by inserting the identity,
%%%%%%%%%%%%%%%%%%\
\begin{equation}
1 \ = \ \frac{1}{5} \, \biggl ( \frac{\partial \ell_{\|\, \mu}}{\partial \ell_{\|\, \mu}} \ + \  \frac{\partial L}{\partial L} \biggr ) \;,
\end{equation}
%%%%%%%%%%%%%%%%%%%%%%
 in (\ref{b5}).
After integrating by parts in the region of convergence,
rewriting the r.h.s in terms of $\mathcal{I}$ from \eq{b5}
and keeping only, potentially, non-vanishing surface terms, 
we arrive at 
%%%%%%%%%%%%%%%%%
\begin{equation}
\mathcal{I} = \frac{i\, \pi^{\omega-2} \, \Gamma(4-\omega)}{4} \oint d^{3} S^{\mu} \,
\frac{ \ell_{\| \, \mu}}{(\ell_{\|}^{2} + \Delta)^{4-\omega}} \  - \
\frac{6\, i\, \pi^{\omega -2 } \Delta}{\Gamma(\omega - 1)} \int d^{4} \ell_{\|}  \int_{0}^{\infty} 
d L \, \frac{L^{2\omega -3}}{(\ell_{\|}^{2} + L^{2} + \Delta)^{4}} \;,
\label{b7}
\end{equation}
%%%%%%%%%%%%%%%
where the first integral is over the Euclidean spatial components of a 4-vector
on a three-sphere. The surface integral converges in $1<\omega < 2$ while
the other in $1<\omega < 3$.  By taking the surface integral on a three-sphere with 
radius $R$ and eventually taking the limit $R\to \infty$ we find
%%%%%%%%%%%%%%%
\begin{equation}
 \oint d^{3} S^{\mu} \,
\frac{ \ell_{\| \, \mu}}{(\ell_{\|}^{2} + \Delta)^{4-\omega}} \  = \ 2  \pi^{2}  \lim_{R\to \infty}
R^{2\omega - 4}\;,
\label{b8}
\end{equation}
%%%%%%%%%%%%%%%
which now  converges in the 
region $\omega \le 2$, that is,  it includes the point $\omega = 2$.
For $\omega <2$ this surface term vanishes while for $\omega = 2$ there
is a finite piece, $2 \pi^{2}$, remaining. This is exactly the term that spoils 
gauge invariance and the equivalence theorem. In DR this term is axiomatically 
absent - allowing the shifting of integral momenta is among DR's main properties.

Turning into the second integral of \eq{b7} we note first that the region of 
convergence includes now $\omega=2$. It gives,
%%%%%%%%%%%%%%%
\begin{equation}
\int d^{4} \ell_{\|}  \int_{0}^{\infty} 
d L \, \frac{L^{2\omega -3}}{(\ell_{\|}^{2} + L^{2} + \Delta)^{4}}  \ = \
\frac{\pi^{2}}{12}\,  \frac{\Gamma(\omega -1) \Gamma(3-\omega)}{\Delta^{3-\omega}}\;.
\label{b9}
\end{equation}
%%%%%%%%%%%%%%%
By placing \eqs{b8}{b9} into \eq{b7}  we finally arrive at \eq{mann2}.
%%%%%%%%%%%%%%%
%\end{appendix}

%%%%%%%%%%% BIBLIOGRAPHY %%%%%%%%%%%%%%%%%
\bibliography{Higgs-Biblio,biblio}{}
\bibliographystyle{utcaps}

\end{document}